\documentclass[12pt]{amsproc}
\usepackage{graphicx}
\begin{document}


	\title{Component Based Programming in Scientific Computing: A Practical Approach/the Viable Approach}
	\author{Zsolt I. L\'az\'ar}
	\address{Department of Theoretical and Computational Physics, Faculty of Physics, Babe\c{s}-Bolyai University, Str. M. Kog\~{a}lniceanu Nr. 1, RO 400084 Cluj-Napoca, Romania}
	\email{zlazar@phys.ubbcluj.ro}
	\thanks{NWO}
	
	\author{Jouke R. Heringa}
	\address{Physical Chemistry \& Molecular Thermodynamics, DelftChemTech,  Delft University of Technology, Julianalaan 136, 2628 BL Delft, The Netherlands}
	\email{J.R.Heringa@tnw.tudelft.nl}
	\thanks{}

%

	\author{Bazil P\^arv}
	\address{Chair of Programming Languages and Methods, Faculty of Mathematics and Computer Science, Babe\c{s}-Bolyai University, Str. M. Kog\~{a}lniceanu Nr. 1, RO 400084 Cluj-Napoca, Romania}
	\email{bparv@cs.ubbcluj.ro}
	\thanks{}

	\author{Simon W. de Leeuw}
	\address{Physical Chemistry \& Molecular Thermodynamics, DelftChemTech,  Delft University of Technology, Julianalaan 136, 2628 BL Delft, The Netherlands}
	\email{S.W.deLeeuw@tnw.tudelft.nl}
	\thanks{}



\date{\today}


\begin{abstract}
Computational scientists are facing a new era where the old ways of developing and reusing code have to be left behind and a few daring steps are to be made towards new horizons. The present work analyzes the needs that drive this change, the factors that contribute to the inertia of the community and slow the transition, the status and perspective of present attempts, the principle, practical and technical problems that are to be addressed in the short and long run.
\end{abstract}

\maketitle


Let us make an imaginary trip into the close future and peek into how you might be doing your research then.
Suppose you are reading a paper presenting an enticing computational study. The attached images are nice but leave certain question marks therefore you decide to be more inquisitive. You type into an application running on your computer the internet address specified by the authors and the next moment you are ready to run exactly the same code that produced the results reported in the paper. The code that you just downloaded, we shall call it a project, is a connected set of computational components with modules each having a well defined purpose, most of them, possibly all, developed by others than the authors of the paper. Using familiar click-and-drag techniques you replace some of these with other similar components provided that they use the same input data and provide the same output variables as the original one. You check whether by replacing any of the numerical components with another, that in principle performs the same tasks, you still reproduce the computations of the article. If you find anything suspect you configure the components differently than the authors suggested and perform some consistency checks. Satisfied or unhappy with what you got you decide to use the same project or parts of it with your data and combine it with your own modules saving months maybe years of work. The missing parts from your project are acquired by simply searching in a practically inexhaustible global repository, dragging into place and connecting computational components developed by a number of independent parties. In case you are new to the field, you select from a large library of pre-made, well-tested domain specific projects that in a few clicks are on your computer ready to be run locally, on your cluster or on the grid resources you have access to. Maybe you have recently written a high-quality computational module that you wish to share with others and obtain the appropriate credits for it. After some effort spent on providing proper documentation you immediately have it released and available for all interested parties irrespective of their programming environment. They can use it with the same ease as you did with the components in the aforementioned project that you have just downloaded. If indeed of good quality, well documented and useful, your component will soon be highly ranked just like some of your papers have a high number of citations. 
Does this sound utopian? Probably. Now, could we try to imagine the opposite of all the above? Unfortunately, there is no need to close our eyes. It is all around us. 

\section*{Crisis in computational science}		\label{S:motivation}

Computational science has recently become aware of a series of limitations that affect its progress. Douglass E. Post calls attention to the different aspects of a crisis \cite{Post04} that bears common traits with the one encountered in software engineering and formulated already in the late sixties \cite{Naur69}. In \cite{Post05} several arguments are put forward regarding the necessity of a change of paradigm especially in  large scale scientific software engineering. We have complemented these ideas relative to the use and development of low and medium scale computational software \cite{Lazar04a} . We point out that a large part of the effort invested in computational research is absorbed by computer related activities such as implementing new code, debugging old code, adapting third party software, setting computing environment, etc. Indeed, due to the rich variety of hardware architectures, operating systems, software libraries, protocols, standards, programming languages, a.s.o, that are in place today, the scientist computer user trying to communicate and share ``business logic" with fellow scientists has to fight the ubiquitous incompatibilities on a day-to-day basis.  Many computational scientists develop their own code mainly because the effort of adapting third party software and the associated long learning curve leaves them without other options. In addition to the low overall efficiency, this state of matters is also responsible for the low quality and unreliability of home-made code, entailing a general lack of trust in third party code. Non-programmers do not find all facilities integrated in one software tool therefore are forced to manually manage the communication of data between a series of applications. This automatically  results in the impossibility for referees and third party scientists to reproduce computer experiments described in scientific publications. In spite of the prevalence of computer aided research there exists no ranking system for scientific software analogous to the number of citations and the (in)famous impact factor assessing one's scientific activity in terms of his/her publications. No wonder that programmer computational scientists do not put much effort into assuring the quality of their code and providing proper documentation for them, hence further undermining the sharing of computational software. 

\section*{The power of sharing}

Even though the concept of sharing rather deserves several volumes than a few lines, a handful of examples are sufficient to realize its outmost importance to human existence. Biological evolution itself, the ultimate wonder of nature, is based on copying information encoded into genes. The sharing of experience and knowledge is maybe the most distinctive feature of the human race allowing  for an evolution beyond the biological one \cite{Blackmore}. The accelerating development of science and technology is the result of ever increasing efficiency in transmitting and sharing information. History demonstrates that a change in information sharing paradigm usually leads to major leaps in the way society works. The early milestones represented by Guttenberg's press, the first phone central and the associated infrastructure, the radio and television were followed by the era of the internet with the World Wide Web as its most prominent component, bringing content sharing to new heights. Peer-to-peer networks, best exemplified by Napster, e-Mule or Kazaa, are another facet of internet based content sharing which accounts for an important part of the total traffic. Grid computing is the next step that most scientists are eager to make. Even though seamless sharing of computer hardware resources is not quite here it is nevertheless expected to open new perspectives especially in computational science. The OpenSource movement with its obvious advantages and in spite of its less obvious limitations provides us with a wealth of high quality software that made its way into most computer systems.

\section*{The promise of component technology}

In the field of computer programming sharing means software reuse. The enabling technique of component software is a hot topic today \cite{Szyperski02}, many viewing it as the holy grail of software reuse. But how important is software reuse in computational science? There are at least three classes of benefits that a mature reuse oriented paradigm would bring about:
\begin{enumerate}
\item {\em Efficiency}  by enhanced reusability, diversity and availability: Scientists would program from scratch only when developing new algorithms and domain specific models. Most of the time they could reuse and integrate seamlessly other's components to support their own research. Finding others' computational works could be done in an instant by selecting from a large distributed repository with a centralized registry employing an advanced classification scheme. Sharing their own code with others can become a standard and effortless procedure. The repository will not only contain powerful, general purpose, elementary components but primarily compound components cleverly built from elementary ones to perform specific tasks of higher complexity. The impact on computer aided education of science is expected to be tremendous. Composed components built from toy and professional, interactive, visualization and other components can be created in the thousands covering all areas of science, all levels of difficulty and most teaching styles. Students will be able to smoothly make the transition from playing with educational projects to doing actual research. 
	
\item {\em Reproducibility} of computer experiments by the author, referees and third party scientists: Computational studies have  tacitly received a similar status as real experiments. Everybody ``experiments" with his/her own tools and methodology that are documented to a certain extent but just for informative purposes. The reader of a paper can seldom perceive a problem with the ``experiment" just by reading the report. Generally, one either accepts the results as they are or tries to create a similar setup on ones own. 
Component software promises all the benefits of genuinely virtual computer experiments that should have been virtual in the first place. As long as the employed components are free from restrictions imposed by copyright issues or if the user can live up to the terms of licenses that apply, the only barrier to reproducing virtual experiments will be a matter of available computing resources. However, in the coming era of grid computing the problem of finite computer resources will manifest in a way that is not obvious at this stage.

\item {\em Quality} due to the strong focus of expert groups and ranking systems. A strong reuse minded computational community comfortably relying on a large repository, will mostly leave the development of new components to specialized groups. This minority even though just a fraction of those programming today will still be large enough to permanently enhance the diversity of the repository. Once reproducing one's result from a computational science paper will be just as straightforward as redoing the calculations from a purely theoretical paper, the system for impact ranking is bound to change. 
Direct evaluation similar to that used for books and movies, evaluation based on the number of downloads or the number of composed components referring to a particular component, specialized repositories associated with scientific journals and magazines and other systems that cannot be anticipated from now are all ways to separate the good from the less good. This natural selection will result in the sedimentation of a layer of high-quality components in all areas of computational science.


\end{enumerate}
Component technology will fully integrate with grid technologies in the years to come confering unprecedented flexibility and efficiency to scientific research.


\section*{State of the art}

The first concrete steps towards component-based scientific computing have been made before the turn of the century. The authors of \cite{Armstrong99}  describe a ``standard for interoperability among high-performance scientific components". They touch upon most fundamental concepts  of component-based programming in the context of high-performance computing  and suggest a standard that they term as ``Common Component Architecture" (CCA). Their recommendation for an interface definition language (IDL) closely follows the CORBA principles. The ideas therein are further developed in \cite{CCA_00}.  This group, however, favors an XML-based language for describing component interfaces. In both cases Java is chosen as the implementation language for the integration framework. Most of later works focus on the integration of the component-based approach into the realm of distributed computing in general and grid computing in particular. Notable efforts have been made by several member institutes of the Common Component Architecture Forum  \cite{CCAF} for delivering different implementations of a CCA framework \cite{CCAFE, SCIRUN2, XCAT}.  Most of the activity is centered around the Babel language interoperability tool \cite{Babel} of the Lawrence Livermoore National laboratory \cite{CASC}. Babel uses the Scientific Interface Definition Language (SIDL) for defining the interfaces of components implemented in any of the programming languages encountered in scientific computing. The palette includes C, Fortran and variants but also higher-level languages such as Java and Python.

Alexandria \cite{Alexandria} is meant to be the future repository for CCA compliant components. As of now it contains no components.

Efforts of more restricted scope can be encountered in different fields of science such as life sciences \cite{IBM_LS}, chemistry \cite{Octet}, nuclear physics \cite{ROOT}, and astrophysics \cite{Flash} to name a few.  The tremendous need in all computational sciences for sharing code is apparent if we consider the popularity of Netlib \cite{NetLib}. Netlib is a collection of mathematical software, papers, and databases with hundreds of numerical libraries available for download. There is an enormous number of over 270 million requests that have been made to the repository. And Netlib only includes mathematical software, no modeling or simulation packages built over them. Unfortunately, there is no uniform way for reusing this large variety of modules originating from different places and being the result of independent and uncorrelated efforts.

\begin{figure}[htb]
\includegraphics[width=15cm]{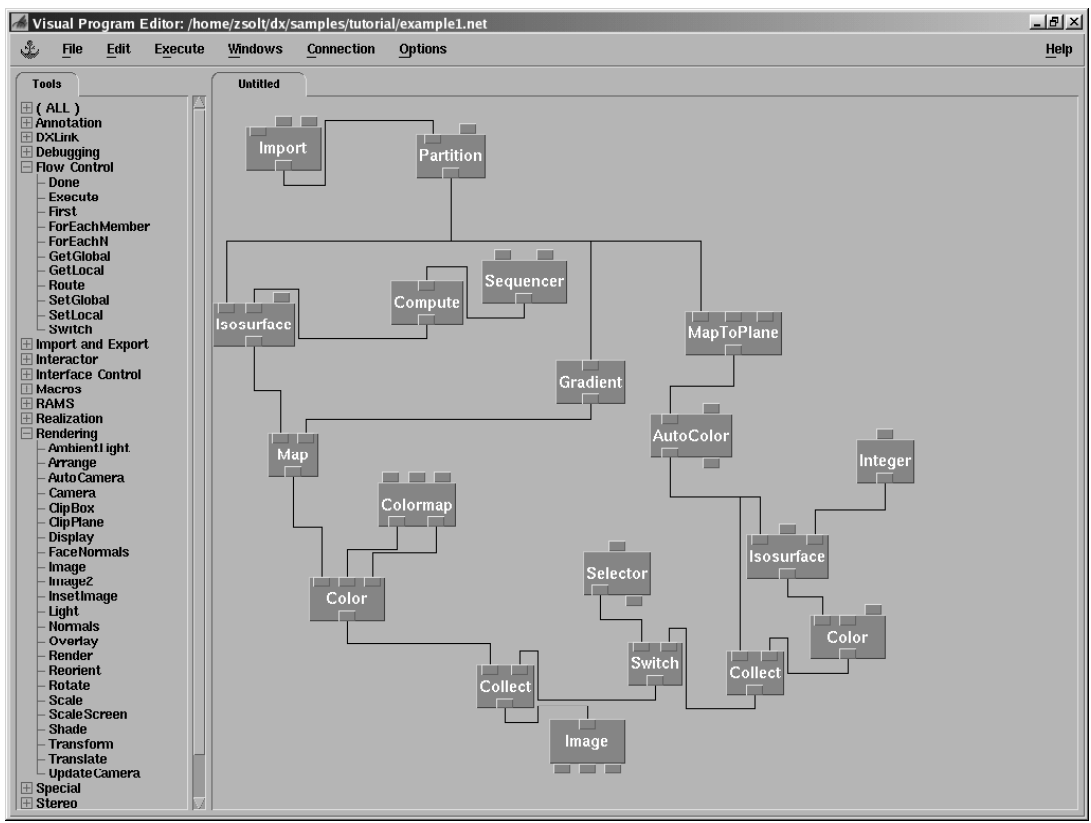}
\caption{\footnotesize Example of flow-based visual programming environment. Representation of a data visualization project in Open Data eXplorer \cite{OpenDX}}\label{fig:opendx}
\end{figure}

The high-level, flow-based, visual programming style aiming at flexibility, efficiency and ease of use is well exemplified by the Open Data eXplorer (OpenDX), a powerful, full-featured software package for the visualization of scientific, engineering and analytical data \cite{OpenDX}. Its open system design is built on familiar standard interface environments. Its sophisticated data model offers great flexibility in creating visualizations by providing hundreds of built-in specialized functions. The user can drag and drop different data filters onto the canvas and connect output to inputs for setting up the data flow diagram (Figure \ref{fig:opendx}). The compatibility of the interfaces is automatically checked. The functions in OpenDX are hard-coded into the application. One can contribute with new components by applying the prescriptions that are set for the interfaces. The new component will become available after the recompilation of the application.

The bottomline is that time is ripe for a substantial change. This is reflected not only in the increasingly pronounced symptoms of an aging paradigm but also in the large number of isolated attempts trying to find a way out. Unfortunately, the present most comprehensive approach promoted by the CCA Forum is plagued by a prohibitively high threshold of adaptation effort which will most probably keep the project in experimental stage for the years to come. 

\section*{The way to go}

The COmputational MODule Integrator (COMODI) \cite{COMODI} initiative emerged from a research collaboration of a more restricted scope involving the Babe\c{s}-Bolyai University in Romania and Delft University of Technology in the Netherlands. It aims at a framework, wiring mechanism, component repository and component developer tools that best fit the needs of computational scientists. At the time of this writing, apart from technical feasibility studies \cite{Lazar04b} and a few prototypes, COMODI basically consists in a set of recommendations and guidelines with regards to the requirements, design, development and dissemination strategies. It is also an invitation of all interested parties for joining in this quest by providing ideas in the first stage, patches, computational components and feedback at the next stage. The COMODI website includes a questionnaire and the reader is very much encouraged to provide a quick and valuable contribution by filling it in.

The essence of the COMODI ``ideology" is that in order to achieve a change of paradigm the employed tools and standards should first adapt to present practices offering a smooth transition to a stage where the scientific community can gradually accommodate to the series of changes ending in a new paradigm. It is based on the realization that shifting to a reuse oriented paradigm is not so much a technical challenge rather than one consisting of a careful requirements analysis, task prioritization and efficient dissemination. 
Based on experience and a preliminary survey we shall rely on the following premises:
\begin{enumerate}
\item computational science is not computer science: the roots of computational science is in natural sciences and engineering not in computer science 
\item computational science demands a new paradigm
\item computational science will not take up any new technologies in the short run unless simpler than those that are in use today
\item the community itself should decide on the new paradigm. A restricted group of promotors can only set the process off and catalyze it.
\item a few, relatively homogeneous groups make up the bulk of the community. The new solution should target these groups instead of trying to be fully comprehensive.
\end{enumerate}

\begin{figure}[htb]
\includegraphics[width=15cm]{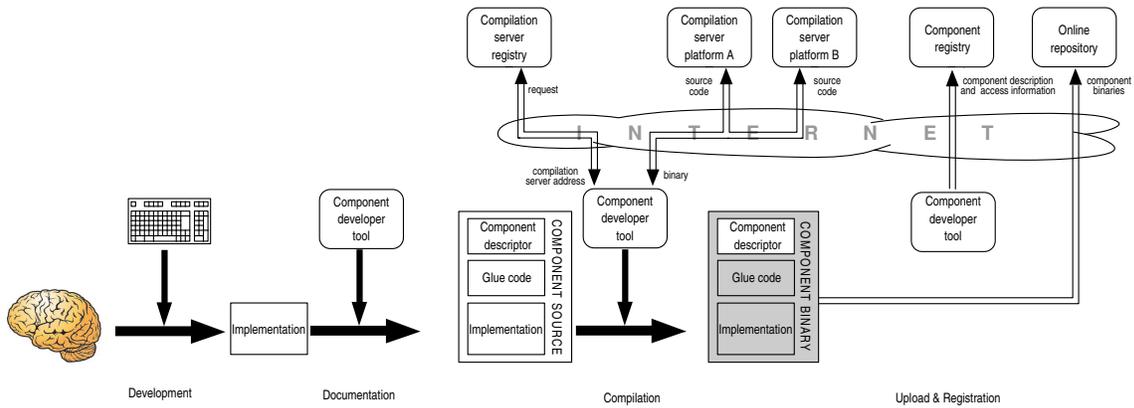}
\caption{\footnotesize Sketch of the four-stage component creation process. In addition to the usual implementation consisting in coding regular computational functions and data structures there are three further stages all assisted by a user friendly developer tool. This tool generates the component descriptor file and glue-code after semi-automatically collecting information pertaining to the behavior, representation, etc. of the component. Then it contacts online compilation servers and returns ready-made binaries for platforms of the developer's choice. The compiled library together with the descriptor file is uploaded by the developer to a place where it can be accessed publicly and the tool registrates the component in the component registry.}
\label{fig:component_development}
\end{figure}

Therefore the ideals that guide the design process of COMODI include:
\begin{enumerate}
\item the envisaged connectivity should not entail significant overhead, i.e., the drop in performance, if not avoidable altogether, should stay within narrow limits. C and Fortran programs built from pieces the usual way by linking together static or shared libraries should not underperform when wrapped as components and linked as a COMODI project.
\item platform independence is one of the most important desiderata. It can be achieved by independent compilation servers that accept source code and returning executables via general web service or other specialized protocols (Figure \ref{fig:component_development}). Component developers will transparently submit compilation requests by a specialized COMODI tool that will mediate the request to compilation servers whose access information is either hard-coded into the application or is obtained from a central registry. Alternatively, component users can request compilation service prior to running their project. Both options exhibit advantages and drawbacks.
\item short learning curve for scientific component developers: the amount of training necessary for moving over to the new technology should be in line with the obtained benefits and with no inhibiting thresholds. This objective assumes a dual view on computational projects. A user friendly graphical environment supporting a high-level, flow-based, visual programming style, such as the Open Data Explorer ( \ref{fig:opendx}), while at the level of individual components maintaining the low-level control computational scientists are used to. 
\item no constructs of high-level abstraction and no new languages required: apart from a few recipe like instructions the computational scientist should not need to deal with new concepts and rules in order to adapt his code to the expectations of the component world.
\item bottom-up construction of the framework: presently, in low level languages like C and Fortran, functions and procedures are the lowest granularity elements of the code. Therefore the design of the framework should be such that this low-level programming style is closely matched within the component oriented visual programming environment offered by the framework
\item no source code modifications required for adapting scientific routines to COMODI: regular procedural code can become connectable with some adjustments. However, the goal is that upon adapting code to COMODI changes in the function interfaces are unnecessary and any connection related extra glue-code should be generated automatically (Figure \ref{fig:component_development}). The actual code can be wrapped as a COMODI component by first documenting its content and describing its dependencies in terms of {\em uses} (input) and {\em provides} (output) ports, adding information concerning its behavior and representation in a COMODI environment, versioning, global naming, etc. The glue code can even be in a separately compiled file containing all the entry points to be used in a COMODI environment while preserving the original code intact ready to be used outside the COMODI environment
\end{enumerate}

For more technical details on how this can be achieved the reader can refer to \cite{Lazar04b}.

The above goals are exclusively of technical nature. However, the COMODI approach distinguishes itself from other attempts by adopting a practical and entrepreneurial view on the objectives. These should manifest themselves in the way the permanently refined requirement analysis is done, the criteria on which design decisions are based and a persistent effort for promoting the solution. In terms of effort-impact considerations we can safely rely on the 80\%-20\% rule which in this case would rather be 90\%-10\%. This means that 10\% of the total effort necessary for a hypothetical complete solution is sufficient for attracting 90\% of the community. The major challenge for covering the targeted 90\% consists in correctly identifying the profile of this 90\% in terms of common practices, needs, tools, etc. and fighting the inertia of a large, conservative community with deeply rooted software development traditions.
Both common sense and experience tell us that such a project should be done in a close collaboration of computational scientists and computer scientists. A major weakness of many present projects attempting to improve computational scientific research is that  they are done without sufficient involvement of the actual target group, the computational scientist. These should define the requirements at each step. The role of computer scientists should be to find the best possible solution in line with the imposed requirements using up-to-date technology. Only a collaboration, that builds on the strong points of all participants, will enable us to build a foundation of a software framework that can assist computational science in its next evolutionary step.


\section*{Acknowledgement}
This work was supported by the Netherland's Organization for Scientific Research (NWO) with grant no. 048.031.003 and by the National University Research Council of Romania (CNCSIS) with grant no. 37/2004. 
\end{document}